\documentstyle[12pt]{article}

\begin{document}

\title{Frame dragging and super--energy}

\author{L. Herrera$^{1}$\thanks{e-mail: laherrera@cantv.net.ve},
J. Carot$^{3}$\thanks{e-mail: jcarot@uib.es},
 and  A. Di Prisco$^{1}$\thanks{e-mail: adiprisc@fisica.ciens.ucv.ve},\\
\small{$^1$Escuela de F\'{\i}sica, Facultad de Ciencias,} \\
\small{Universidad Central de Venezuela, Caracas, Venezuela.}\\
\small{$^3$Departament de  F\'{\i}sica,}\\
\small{Universitat Illes Balears, E-07122 Palma de Mallorca, Spain}\\
} \maketitle
\begin{abstract}

\end{abstract}
We show that the vorticity appearing in stationary vacuum spacetimes
is always related to the existence of a flow of super--energy on the
plane orthogonal to the vorticity vector. This result, toghether
with the previously established link between vorticity and
super--energy in radiative (Bondi--Sachs) spacetimes  strength
further the case for this latter quantity as the cause of frame
dragging.

\section{Introduction}
The appearance of vorticity in the congruence of world--lines of
observers in the gravitational field of a massive rotating ball
(Lense--Thirring effect), led Shiff \cite{Shiff} to propose the use
of gyroscopes to measure such an effect. Since then this idea has
been developed extensively (see \cite{Rindler}--\cite{ Iorio1})  and references cited
therein).

However, the appearance of vorticity is not always related (at least
explicitly) to rotating sources. Indeed, this is the case of the
field of a magnetic dipole plus an electric charge \cite{Bonnor},
and the case of gravitational radiation \cite{HHP}--\cite{HSC}.

In the former case the vorticity is  accounted for by the existence
of a flow of electromagnetic energy on the plane orthogonal to the
vorticity vector. As a matter of fact it appears that   in all
stationary electrovacuum solutions \cite{colombia}, at least part of
the vorticity has that origin.

In the case of gravitational radiation (Bondi--Sachs) we have
recently shown \cite{sachs} that  the appearing vorticity is
related to the existence of a flow of super--energy on the plane
orthogonal to the vorticity vector.

Here we prove that  such a link between a flow of super--energy and
vorticity, is also present in stationary vacuum spacetimes.

In order to motivate our study, we shall start by considering simple particular cases, before dealing with the general stationary situation. Thus, we shall first present the
Lense--Thirring case (Kerr up to the first order in $m/r$ and
$a/r$), then we will treat the Kerr case, and finally we establish 
the aforementioned link for the general stationary vacuum spacetime.

Doing so we provide an ``universal'' mechanism (i.e. one which applies to all known situations) for the occurrence of frame dragging. At the same time this result brings out the relevance of the Bel--Robinson tensor in the study of self--gravitating systems.
\section{The Lense--Thirring metric }
As is well known the Lense--Thirring metric \cite{lense}
\begin{eqnarray}
ds^2&=&-\left(1-\frac{2m}{r}\right)dt^2+\left(1+\frac{2m}{r}\right)\left(dr^2+r^2d\theta^2+r^2\sin^2\theta d\phi^2\right)\nonumber\\
&+&\frac{4J\sin^2\theta}{r}d\phi dt \label{LT1}
\end{eqnarray}
describes the gravitational field outside a spinning sphere of
constant density, and is valid up to first order in $m/r$ and
$J/r^2$, with $m$ and $J$ denoting the mass and the angular momentum
respectively.

It is also well known that, up to that  order, it is the Kerr
metric, with the identification
\begin{equation}
ma=-J \label{ltk}
\end{equation}
where $a$ is the Kerr parameter \cite{kerr}

Now, the congruence of the world--line of observers at rest in the
frame of (\ref{LT1}) is
\begin{equation}
u^{\alpha}=\left(\frac{1}{\sqrt{1-\frac{2m}{r}}},0,0,0\right),
\label{LTu}
\end{equation}

and the vorticity vector, defined as usual by
\begin{equation}
\omega^\alpha=\frac{1}{2}\eta^{\alpha\eta\iota\lambda}u_\eta
u_{\iota,\lambda},\label{vv}
\end{equation}
yields, up to order $a/r$ and $m/r$
\begin{equation}
\omega^r=\frac{2ma\cos\theta}{r^3} \label{w1}
\end{equation}
\begin{equation}
\omega^\theta=\frac{ma\sin\theta}{r^4} \label{w2}
\end{equation}
or, for the absolute value of the vector $\omega^\alpha$
\begin{equation}
\Omega=(\omega^\alpha
\omega_{\alpha})^{1/2}=\frac{ma}{r^3}\sqrt{1+3\cos^2\theta}
\label{LT2}
\end{equation}

At the equator ($\theta=\frac{\pi}{2}$)
\begin{equation}
\Omega=\frac{ma}{r^3} \label{LT3}
\end{equation}
which is a very well known result.

The leading term of the super--Poynting gravitational vector at the equator, calculated from
(\ref{LT1}), is (see next section), 
\begin{equation}
P^\Phi \approx 9\frac{m^2}{r^2} \frac{a}{r}\frac{1}{r^5}\label{LT3bis}
\end{equation}
implying that $P^\phi=0  \Leftrightarrow a=0
\Leftrightarrow\omega^{\alpha} = 0$.

We shall now prove   our conjecture  in the exact case (Kerr metric).
\section{The Kerr metric}
In Boyer--Linquist coordinates, the Kerr metric takes the form
\begin{eqnarray}
ds^2 &=& \left(-1 + \frac{2mr}{r^2 + a^2 \cos^2{\theta}}\right) dt^2 - \left(\frac{4 m a r \sin^2{\theta}}{r^2 + a^2 \cos^2{\theta}}\right) dt d\phi \nonumber \\
&+& \left(\frac{r^2 + a^2 \cos^2{\theta}}{r^2 - 2mr + a^2}\right) dr^2 + \left(r^2 + a^2 \cos^2{\theta}\right) d\theta^2 \nonumber \\
&+& \left(r^2 \sin^2{\theta} + a^2 \sin^2{\theta} + \frac{2 m r a^2
\sin^4{\theta}}{r^2 + a^2 \cos^2{\theta}}\right) d\phi^2
\label{kerr1}
\end{eqnarray}
The congruence of world--line of observers at rest in (\ref{kerr1})
is defined by the vector field
\begin{equation}
u^{\alpha}=\left(\frac{1}{\sqrt{1-\frac{2mr}{r^2+a^2
\cos^2\theta}}},0,0,0 \right) \label{kerr2}
\end{equation}
There are two non--vanishing components of the vorticity vector for
such a congruence, which are
\begin{eqnarray}
\omega^r&=& 2 m r a \cos{\theta} \left(r^2 - 2mr + a^2\right) \left(r^2 + a^2 \cos^2 {\theta}\right)^{-2} \nonumber \\
&&\left(r^2 - 2mr + a^2 \cos^2{\theta}\right)^{-1} \label{kerr3}
\end{eqnarray}
and
\begin{eqnarray}
\omega^\theta&=& m a \sin{\theta} \left(r^2 - a^2\cos^2{\theta}\right) \left(r^2 + a^2 \cos^2{\theta}\right)^{-2} \nonumber \\
&&\left(r^2 - 2mr + a^2 \cos^2{\theta}\right)^{-1} \label{kerr4}
\end{eqnarray}
which of course coincide with (\ref{w1}) and (\ref{w2}) up to first
order in $m/r$ and $a/r$.

Next, the  super--Poynting vector  based on the Bel--Robinson
\cite{Bel} tensor, as defined in Maartens and Basset
\cite{maartens}, is
\begin{equation}
P_{\alpha}=\eta_{\alpha \beta \gamma
\delta}E^{\beta}_{\rho}H^{\gamma \rho}u^{\delta}, \label{p1}
\end{equation}

where $E_{\mu\nu}$ and $H_{\mu\nu}$, are the electric and magnetic
parts of Weyl tensor, respectively, formed from Weyl tensor
$C_{\alpha \beta \gamma \delta}$ and its dual $\tilde C_{\alpha
\beta \gamma \delta}$ by contraction with the four velocity vector,
given by
\begin{equation}
E_{\alpha \beta}=C_{\alpha \gamma \beta \delta}u^{\gamma}u^{\delta}
\label{electric}
\end{equation}
\begin{equation}
H_{\alpha \beta}=\tilde C_{\alpha \gamma \beta
\delta}u^{\gamma}u^{\delta}= \frac{1}{2}\eta_{\alpha \gamma \epsilon
\delta} C^{\epsilon \delta}_{\quad \beta \rho} u^{\gamma} u^{\rho},
\label{magnetic}
\end{equation}

Then, a direct calculation of $P^\mu$ using the package GR--Tensor
running on Maple yields, for the Kerr space-time:
\begin{equation}\label{SPkerr}
    P^\mu = \left(P^t,P^\phi,0,0\right)
\end{equation}
where
\begin{eqnarray}
P^{t}&=&-18 m^3 r a^2 \sin^2{\theta}\left(r^2 - 2mr + a^2\sin^2{\theta} + a^2 \right) \left(r^2 + a^2\cos^2{\theta}\right)^{-4} \nonumber\\
&&  \left(r^2 - 2 m r + a^2\cos^2{\theta} \right)^{-2}
\left(\frac{r^2 -2 m r + a^2\cos^2{\theta}}{r^2 +
a^2\cos^2{\theta}}\right)^{-1/2} \label{pkt}
\end{eqnarray}
and
\begin{eqnarray}
P^{\phi}&=&9 m^2 a \left(r^2 - 2mr + 2 a^2 - a^2 \cos^2{\theta}\right) \left(\frac{r^2 - 2mr + a^2 \cos^2{\theta}}{r^2 + a^2 \cos^2{\theta}}\right)^{-1/2}\nonumber \\
&& \left(r^2 + a^2 \cos^2{\theta}\right)^{-4} \left(r^2 - 2mr + a^2
\cos^2{\theta}\right)^{-1} \label{pkfi}
\end{eqnarray}
From the above it can be seen that $P^\phi=0  \Leftrightarrow a=0
\Leftrightarrow\omega^{\alpha} = 0$. In other words: there is always
an azimuthal flow of super--energy, as long as $a\neq 0$, and
viceversa, the vanishing of such a flow, implies $a=0$. Also observe
that the leading term in power series of $m/r$ and $a/r$ in
(\ref{SPkerr}) is of order $(m/r)^2$.

Let us now consider the general stationary and axisymmetric vacuum
case.

\section{The general case}
The line element for a general stationary and axisymmetric vacuum
spacetime may be written as \cite{kramer}
\begin{equation}
ds^2=-fdt^2+2f\omega dt d\phi+f^{-1} e^{2
\gamma}(d\rho^2+dz^2)+(f^{-1}\rho^2-f\omega^2)d\phi^2
\label{metrica}
\end{equation}
where $x^0=t$; $x^1=\rho$; $x^2=z$ and $x^3=\phi$ and metric
functions depend only on $\rho$ and $z$ which must satisfy the
vacuum field equations:
\begin{equation}\label{EFE1}
    \gamma_\rho = \frac{1}{4\rho f^2}\left[ \rho^2\left( f_\rho^2-f_z^2  \right)-f^4 \left( \omega_\rho^2-\omega_z^2  \right)   \right]
\end{equation}

\begin{equation}\label{EFE2}
    \gamma_z = \frac{1}{2\rho f^2}\left[ \rho^2 f_\rho f_z - f^4 \omega_\rho\omega_z   \right]
\end{equation}

\begin{equation}\label{EFE3}
    f_{\rho \rho} = - f_{zz} - \frac{f_\rho}{\rho} -\frac{f^3}{\rho^2}\left( \omega_\rho^2+\omega_z^2 \right) + \frac{1}{f} \left(  f_\rho^2 +
    f_z^2\right)
\end{equation}

\begin{equation}\label{EFE4}
    \omega_{\rho \rho} = - \omega_{zz} + \frac{\omega_\rho}{\rho} -\frac{2}{f}\left(f_\rho \omega_\rho + f_z\omega_z \right)
\end{equation}

The four velocity vector for an observer at rest in the frame of
(\ref{metrica}) is
\begin{equation}
u^{\alpha}=(f^{-1/2},0,0,0) \label{four}
\end{equation}

The super-Poynting vector can now  be calculated for the general
class of spacetimes represented by the above metric (\ref{metrica}),
(i.e.: without making any assumption about the matter content of the
spacetime), and one gets (using again GR--Tensor):
\begin{equation}\label{SP0}
    P^\mu = (P^t,P^\phi,0,0) \quad {\mathrm{with}} \quad P^t = \omega
    P^\phi, \quad {\mathrm{hence}}\quad  P_\mu = (0,\frac{\rho^2}{f}P^\phi,0,0)
\end{equation}

Thus, the relevant quantity is $P^\phi$ which is given by (again in
the general case, i.e.: without taking into account the field
equations):
\begin{eqnarray}
  \nonumber P^\phi = f^{3/2}e^{-4\gamma}\rho^{-5} \left\{  A11  \right\}  \\
\label{P20}
\end{eqnarray}

Substituting now the vacuum field equations (\ref{EFE1}-\ref{EFE4})
in the above expression one gets:

\begin{eqnarray}
  \nonumber P^\phi = -\frac{1}{32}f^{-3/2}e^{-4\gamma}\rho^{-5} \left\{  A12  \right\}  \\
\label{P20}
\end{eqnarray}
where $A11$ and $A12$ are given in the Appendix.

Now, it has been shown in \cite{colombia} that for the general
metric (\ref{metrica}) the following relations hold
\begin{equation} \label{ultima}
H_{\alpha \beta} = 0 \Leftrightarrow \omega^{\alpha} = 0
\Leftrightarrow \omega =0 .
\end{equation}
and of course we know that
\begin{equation} \label{ultima}
H_{\alpha \beta} = 0 \Rightarrow P^\mu = 0\, .
\end{equation}
So,  what we want to show here is  that $P^{\mu} =0$ implies
necessarily that $\omega =0$  the solution becomes then static, not
just stationary, and therefore the so-called ``dragging of inertial
frames'' effect disappears as the vorticity vanishes.

In other words we want to establish the relation

\begin{equation} \label{ultimaI}
 P^\mu = 0 \Leftrightarrow H_{\alpha \beta} = 0 \Leftrightarrow \omega^{\alpha} = 0 \Leftrightarrow \omega =0\, .
\end{equation}
Now, as the above equation is far too complicated to be treated in
full, we shall start instead by analyzying what happens in the
neighborhood of the symmetry axis $\rho = 0$ and far away from any
matter source along it, that is, for $z \rightarrow \infty$. In so
doing, the following two assumptions will be made:
\begin{enumerate}
  \item The spacetime is regular at the axis.
  \item The spacetime is asymptotically flat in spacelike
  directions.
\end{enumerate}

\subsection{The super-Poynting vector in a neighborhood of the axis}

The geometry of axisymmetric spacetimes in the vicinity of the axis
was studied in \cite{Carot2000} (see also \cite{Carlson1980}
and\cite{kramer}). It then follows that $g_{t\phi}$ must tend to
zero when $\rho \rightarrow 0$ at least as $\rho^2$, $g_{\phi\phi}$
tends to zero as $\rho^2$, and $g_{t t}, g_{\rho\rho}$ and $g_{zz}$
cannot vanish on the symmetry axis. All these imply (but are not
equivalent to) the so-called `elementary flatness condition' on the
axis, namely:
\begin{equation}
    \frac{X^a X_a}{4X}  \longrightarrow
    1 \qquad  X \equiv g_{\phi \phi} \nonumber
\end{equation}

Let us then put
\begin{equation}\label{funciones}
    \omega (\rho,z) = \rho^{2+k} A(z) + O(\rho^3), \qquad f(\rho,z) = m(z) + \rho n(z) +
    \rho^2 S(z) + O(\rho^3)
\end{equation}
where $k \geq 0$ is a constant and $m(z)\neq 0$ necessarily;
further, the elementary flatness condition mentioned above implies
that $\gamma \rightarrow 0$ as $\rho$ tends to zero, and expanding
$P^\phi$ in a power series around $\rho = 0$ one gets (in the case
$k=0$):

\begin{eqnarray}
    P^\phi \propto -\frac{8Am^2n^2}{\rho^2} -
    \frac{8m^2}{\rho}\left\{-4 {A}^{3}n{m}^{4}+3 {m}^{2}n_{{z}}A_{{z}}
    +{m}^{2}nA_{{z,z}} + 4mA_{
{z}}nm_{{z}} \right. \nonumber\\
 \left. -4 m A n S-2 mAm_{{z,z}}n+6
mAn_{{z}}m_{{z}}+3 A{m_{{z}}} ^{2}n-A{n}^{3} \right\} \nonumber \\ +
2\,m \left\{ -16\,{A}^{5}{m}^{8}   + \left(
8\,{A}^{2}A_{{z,z}}+12\,{A_{{z }}}^{2}A \right) {m}^{6}+ \right.
\nonumber
\\ \left( -16\,{A}^{3}m_{{z,z}}+56\,A_{{z}}{A}^
{2}m_{{z}}-32\,{A}^{3}S \right) {m}^{5} + \left(
-72\,{A}^{3}{n}^{2}+24 \,{A}^{3}{m_{{z}}}^{2} \right) {m}^{4}
 \nonumber \\ + \left( 8\,SA_{{z,z}}+24\,A_{{z
}}S_{{z}}+8\,m_{{z,z}}A_{{z,z}} \right) {m}^{3}+ \left[
10\,{n}^{2}A_{ {z,z}}+ \left(
76\,A_{{z}}n_{{z}}-8\,An_{{z,z}} \right) n \right. \nonumber \\
\left. + 24\,A{n_{{z}}}^{2} +
20\,A_{{z}}m_{{z,z}}m_{{z}}-16\,A{S}^{2} \right. \nonumber\\
\left.
-16\,Am_{{z,z}}S-2\,A_{{z,z}}{m_{{z}}}^{2}+32\,A_{{z}}Sm_{{z}}
+48\,A S_{{z}}m_{{z}} \right] {m }^{2} \nonumber \\ + \left[  \left(
-12\,AS-20\,Am_{{z,z}}+28\,A_{{z}}m_{{z}}
 \right) {n}^{2} \right. \nonumber \\  \left. +  88\,An_{{z}}nm_{{z}}-2\,A_{{z}}{m_{{z}}}^{3}+24\,A{m_
{{z}}}^{2}S-4\,Am_{{z,z}}{m_{{z}}}^{2} \right] m \nonumber
\\ \left. +7\,A{m_{{z}}}^{4}+10 \,A{m_{{z}}}^{2}{n}^{2}-A{n}^{4} \right\}  + O(\rho) \qquad \qquad \label{P2eje}
\end{eqnarray}

Setting $P^\phi=0$ all over the spacetime implies that the above
must also vanish in a neighborhood of $\rho = 0$, which in turn
implies that the coefficients of $\rho^{-2}$, $\rho^{-1}$ and
$\rho^{0}$ must be zero. Since $m\neq 0$, it must be either $A=0$
(which is what we aim at showing), or else $n=0$.

Let us assume that $n=0$, the coefficients of the terms in
$\rho^{-2}$ and $\rho^{-1}$ then vanish identically, whereas the
last term is much reduced.

Further, our second requirement above (asymptotic flatness), implies
that far away from the source and in a small neighborhood of the
axis, $m(z) = 1$, $m_z, m_{zz} =0$, and also $S(z) =0$ (for
otherwise $f \neq 1$ at infinity in spatial directions), therefore
one is left with:
\begin{equation}\label{edo1}
    2AA_{zz} + 3A_{z}^2 -4A^4 = 0
\end{equation}
which must hold for $z$ large and in a neighborhood of the axis.
Further, its solution must be bounded, since otherwise $\omega$
would increase without limit thus violating again the condition of
asymptotic flatness.

This is an autonomous equation, a first integral of which can be
readily found to be:
\begin{equation}\label{edo2}
     A_{z} =  \pm \frac{1}{\sqrt{7}} \frac{1}{A^2}\sqrt{A(4A^7 +
     C)}
\end{equation}
$C$ being a constant of integration. It is then immediate to show
using numerical simulations that the function $A$ diverges for large
values of $z$, and therefore it must vanish, which is what we wanted
to show.

So far, we have only analyzed the case $k=0$, however it is a simple
matter to check that field equation (\ref{EFE4}) together with
(\ref{funciones}) rule out all values of $k>0$.

In order to complete our proof we have to show that the result above
(the vanishing of vorticity, implied by the vanishing of
super--Poynting, within an  infinite cylinder around the axis of
symmetry), can be analytically extended to the whole spacetime.

In other words we have to prove that it is not possible to smoothly
match an static axially symmetric, asymptotically flat vacuum
spacetime to a stationary (non-static) axially symmetric vacuum
spacetime which is also asymptotically flat, across an infinite
cylinder (say $\Sigma$) around the axis of symmetry.

Such a matching is not possible \cite{Mars}, but it can also   be
checked at once from the continuity of the first fundamental forms
on the cylinder. Indeed, this last requirement implies that $\omega$
should vanish also at the outer part of $\Sigma$, thereby indicating
that the static condition can be analytically extended.
\section{Conclusions}
We have seen that Bonnor's original idea to associate the frame
dragging in some electrovac solutions with the existence of a flow
of electromagnetic energy (as described by the Poynting vector), can
be successfully extended to the same effect in vacuum stationary
spacetimes, by replacing the flow of electromagnetic energy by a
flow of super--energy, as described by the super--Poynting vector
defined from the Bel--Robinson tensor.

Due to the lack of a covariant definition of gravitational energy, super--energy appears to be the best candidate for
playing such a role. On the other hand, the fact that it is
also associated to frame dragging  in radiative spacetimes,
reinforces further our conjecture, that it is responsible for such
an effect in any general relativistic scenario.

Before concluding, the following remark is in order:  All along this work,  the vorticity is calculated for a congruence  with a distinct physical meaning, namely the congruence of  worldlines of observers at rest  with  respect  to the source, i.e. observers at rest in the frame of (\ref{LT1}),   (\ref{kerr1}) and  (\ref{metrica}), respectively.This is particularly clear  in the case of the Lense--Thirring  metric (\ref{LT1}). 

\section*{Appendix}

\begin{eqnarray}
A11 &=& \left[-2\rho \omega_\rho(\omega_\rho^2 +
\omega_z^2)\gamma_\rho -2 \rho\omega_z (\omega_\rho^2 +
\omega_z^2)\gamma_z -\omega_z^2\omega_{\rho\rho}\rho \right.
\nonumber\\
&&\left.+ \omega_z^2\omega_{zz}\rho +
4\omega_z\omega_\rho\omega_{\rho z}\rho - \omega_\rho^2
\omega_{zz}\rho - \omega_\rho^3 +
\omega_\rho^2\omega_{\rho\rho}\rho\right]f^4
\nonumber\\
&&+3\rho(\omega_\rho^2 + \omega_z^2)(\omega_zf_z + \omega_\rho
f_\rho)f^3 -2\rho (-2\rho\gamma_z\omega_{\rho z}+ 2\gamma_z^2
\omega_\rho \rho
\nonumber\\
&&+2\rho\gamma_\rho^2\omega_\rho + \gamma_z\omega_z +
\gamma_\rho\omega_\rho + \rho\gamma_\rho\omega_{zz} - \rho
\gamma_\rho\omega_{\rho\rho}) f^2
\nonumber\\
&&+\left[4\rho^3(f_z\omega_z + f_\rho\omega_\rho)\gamma_\rho^2 +
2\rho^2 (\rho f_{zz} \omega_\rho - 2\rho f_{\rho z} \omega_z -
2f_z\omega_z \right.
\nonumber\\
&&\left. + 4f_\rho \omega_\rho - 2\rho f_z \omega_{\rho z} - \rho
f_{\rho\rho}\omega_\rho - \rho f_\rho \omega_{\rho\rho} + \rho
f_\rho \omega_{zz})\gamma_\rho \right.
\nonumber\\
&&\left. + 4\rho^3(f_z\omega_z + f_\rho \omega_\rho)\gamma_z^2 +
2\rho^2(4f_\rho \omega_z + \rho f_z \omega_{\rho\rho} - 2\rho
f_{\rho z} \omega_\rho \right.
\nonumber\\
&& \left. + \rho f_{\rho\rho}\omega_z - 2\rho f_\rho \omega_{\rho z}
- \rho f_{zz} \omega_z - \rho f_z \omega_{zz} + 2\omega_\rho
f_z)\gamma_z\right.
\nonumber\\
&& \left. +4\rho^3 f_{\rho z} \omega_{\rho z} - \rho^3 f_{zz}
\omega_{\rho\rho} - \rho^3 f_{\rho\rho}\omega_{zz} + \rho^2 f_{zz}
\omega_\rho - 2 \rho^2 f_{\rho z} \omega_z\right.
\nonumber\\
&& \left. -\rho^2 f_{\rho\rho} \omega_\rho + \rho^3 f_{zz}
\omega_{zz} + \rho^3 f_{\rho\rho} \omega_{\rho\rho}\right] f -
6\rho^3(f_\rho^2 + f_z^2)\omega_\rho\gamma_\rho
\nonumber \\
&& -6\rho^3(f_\rho^2 + f_z^2)\omega_z \gamma_z + 3 \rho^3
(f_{\rho\rho} f_\rho \omega_\rho + f_{zz} f_z \omega_z + 2f_{\rho z}
f_z \omega_\rho
\nonumber\\
&&- f_{\rho\rho} f_z \omega_z +2 f_{\rho z} f_\rho \omega_z - f_{zz}
f_\rho \omega_\rho) \label{A11}
\end{eqnarray}

\begin{eqnarray}
A12 &=& \omega_\rho (-7 \omega_z^4- 6 \omega_\rho^2 \omega_z^2 +
\omega_\rho^4) f^9
\nonumber\\
&& + \left[-\rho \omega_\rho f_\rho ( \omega_z^4 + \omega_\rho^4 + 2
\omega_\rho^2 \omega_z^2 ) - \rho f_z \omega_z (\omega_\rho^4 + 2
\omega_z^2 \omega_\rho^2 +\omega_z^4) \right]f^8
\nonumber \\
&& + \left[-4 \rho \omega_{zz}(\omega_\rho^2 + 3\omega_z^2) + 4
\omega_\rho (-2 \rho \omega_z \omega_{\rho z} + \omega_z^2)\right]
f^7
\nonumber\\
&& + \left[ 4 \rho \omega_z(-8 \omega_z^2 f_z + \rho \omega_z^2
f_{\rho z} -3 \rho \omega_\rho \omega_z f_{zz} -3 \rho \omega_\rho^2
f_{\rho z}-\rho f_\rho \omega_z \omega_{zz} \right.
\nonumber\\
&& \left.  - 5 \omega_\rho^2 f_z - 2 \rho f_\rho \omega_\rho
\omega_{\rho z} -2 \omega_z \omega_\rho f_\rho - 2 \rho f_z
\omega_\rho \omega_{zz} +\rho f_z \omega_z \omega_{\rho z} )\right.
\nonumber\\
&& \left. +4 \rho \omega_\rho (-\rho \omega_\rho f_z \omega_{\rho z}
+ \rho \omega_\rho f_\rho \omega_{zz} + \omega_\rho^2 f_\rho + \rho
\omega_\rho^2 f_{zz}) \right]f^6
\nonumber\\
&& + \left[-6 \rho^2 \omega_\rho^3 (f_z^2 + f_\rho^2) -2 \rho^2 f_z
\omega_\rho \omega_z(2\omega_\rho f_\rho + 5 \omega_z f_z) \right.
\nonumber\\
&& \left. + 2 \rho^2 f_\rho \omega_z^2 (2 \omega_z f_z - \omega_\rho
f_\rho) \right]f^5 + \left[8\rho^2(f_{\rho z} \omega_z - f_\rho
\omega_{zz})\right.
\nonumber\\
&& \left. -16\rho^3(f_{\rho z} \omega_{\rho z} + f_{zz} \omega_{zz})
+ 10 \rho^3 f_z \omega_z f_\rho \omega_\rho (f_z \omega_z + f_\rho
\omega_\rho)\right.
\nonumber\\
&& \left. - 2\rho^3 f_z(f_\rho f_z \omega_\rho^3 + f_\rho^2
\omega_z^3) +2 \rho^3 f_\rho^3(\omega_\rho^3 - \omega_\rho
\omega_z^2)\right.
\nonumber\\
&& \left. +2 \rho^3 f_z^3 (\omega_z^3 -\omega_z \omega_\rho^2)
\right] f^4 + \left[-24\rho^3 f_{\rho z}( \omega_\rho f_z + \omega_z
f_\rho)\right.
\nonumber\\
&& \left. + 4 \rho^2 f_\rho (f_\rho \omega_\rho -4 f_z \omega_z) +4
\rho^3 (3 f_\rho^2 \omega_{zz} +2 f_{zz} f_\rho \omega_\rho -2
\omega_{\rho z} f_\rho f_z\right.
\nonumber\\
&&\left. + f_z^2 \omega_{zz} -10 f_{zz} \omega_z f_z) \right] f^3 +
\left[4 \rho^4 f_{\rho z}( f_\rho^2 \omega_z -f_z^2 \omega_z +2
\omega_\rho f_\rho f_z) \right.
\nonumber\\
&&\left. + 4 \rho^4f_{zz} (f_z^2 \omega_\rho - f_\rho^2 \omega_\rho
+ 2 \omega_z f_\rho f_z ) + 4 \rho^4 \omega_{\rho z} (-f_z^3 + 3f_z
f_\rho^2)\right.
\nonumber\\
&& \left. +4 \rho^4 \omega_{zz} (-f_\rho^3 + 3 f_\rho f_z^2) + 4
\rho^3 (4 f_\rho^2 f_z \omega_z - 3 f_z^2 f_\rho \omega_\rho \right.
\nonumber\\
&& \left. - 2 f_\rho^3 \omega_\rho + 3 f_z^3 \omega_z) \right] f^2 +
\left[\rho^4 \omega_\rho (14 f_z^2 f_\rho^2 - 7f_z^4 +5f_\rho^4)
\right.
\nonumber\\
&& \left.+ 4 \rho^4 \omega_z(f_\rho^3 f_z + 5 f_z^3 f_\rho)\right]f
-\rho^5 \omega_z(2f_z^3 f_\rho^2+ f_z^5 + f_z f_\rho^4 )
\nonumber\\
&& - \rho^5 \omega_\rho (2 f_\rho^3 f_z^2 + f_\rho^5 + f_\rho f_z^4 )
\label{A12}
\end{eqnarray}

\section*{Acknowledgements}
One of us (JC) gratefully acknowledges financial support from the
Spanish Ministerio de Educaci\'{o}n y Ciencia through the grant
FPA2004-03666. LH wishes to thank  Universitat  Illes Balears for financial support and hospitality. ADP also acknowledges hospitality of the
Physics Department of the  Universitat  Illes Balears. LH and ADP acknowledge financial support from the CDCH at Universidad Central de Venezuela under grant PI 03.11.4180.1998.

\end{document}